\documentclass[12pt]{article} 

\def\bs{\bigskip}

\begin{document}

\baselineskip=12pt plus 1pt minus 1pt 

\begin{center}

{\large \bf Deformed Harmonic Oscillators for Metal Clusters and 
Balian--Bloch Theory} 

\bigskip\bigskip

{Dennis Bonatsos$^*$, D. Lenis$^*$, P. P. Raychev$^\dagger$, P. A. 
Terziev$^\dagger$} 

\bs 
{$^*$ Institute of Nuclear Physics, N.C.S.R. ``Demokritos''}

{GR-15310 Aghia Paraskevi, Attiki, Greece}

\bs 
{$^\dagger$ Institute for Nuclear Research and Nuclear Energy, Bulgarian 
Academy of Sciences}

{72 Tzarigrad Road, BG-1784 Sofia, Bulgaria}

\bs

{\bf Abstract}

\end{center}

The predictions for the shell structure of metal clusters of the
three-dimensional $q$-deformed harmonic oscillator (3D $q$-HO),  utilizing
techniques of quantum groups and having the symmetry
u$_q$(3)$\supset$so$_q$(3), are compared to the restrictions imposed by
the periodic orbit theory of Balian and Bloch,
of electrons moving in a spherical cavity. It is shown that agreement
between the predictions of the two models is established  through the
introduction of an additional term to the Hamiltonian of the 3D $q$-HO,
which does not influence the predictions for supershells.
This term preserves the u$_q$(3)$\supset$so$_q$(3) symmetry, while
in addition it can be derived through a variational procedure, analogous
to the one leading from the usual harmonic oscillator to the Morse
oscillator by introducing the concept of the Variable Frequency Oscillator
(VFO).

\bs\bs 

\section{Introduction} %1

Algebraic models are popular in several branches of physics \cite{Iac1,Iac2}.
In ideal cases, algebraic models are able to describe in good approximation
several properties of a physical system using a limited number
of appropriate (usually collective) degrees of freedom. In addition
to providing successful predictions by themselves, exactly soluble algebraic
models can also serve as a useful testground for more sophisticated
microscopic theories, in which heavy numerical work is inevitable.

An algebraic approach to the structure of metal clusters
\cite{deHeer611,Brack677} has been introduced
recently \cite{PRA62,IJQC89,PRA65,AQC40,IJQC377}, taking advantage of the
three-dimensional
$q$-deformed harmonic oscillator (3D $q$-HO) \cite{Terziev},
which is constructed using the techniques
of quantum algebras (quantum groups) [see \cite{PPNP} and references therein].
The symmetry of this oscillator
is u$_q$(3)$\supset$so$_q$(3) \cite{STK437,VdJL213,VdJ1799,Quesne81,JPA29},
while its derivation involves use
of irreducible tensor operators under so$_q$(3) \cite{Quesne81,STK593,STK690}.
In terms of the 3D $q$-HO
a good description of the magic numbers of alkali clusters (up to
1500 atoms, which is the limit of validity for theories based on the
filling of electronic shells \cite{Martin209,Clem44}), as well as of Al
clusters,
has been obtained \cite{PRA62}.
In addition, it has been proved that supershells occur naturally  in this
model \cite{IJQC89,PRA65},
which is characterized by only one parameter (the deformation
parameter $q=e^\tau$, with $\tau$ being real) in addition to the overall
scale.

An alternative approach to shell structure coming from mathematical physics
has been introduced long before the discovery of metal clusters
\cite{Knight2141} by Balian and Bloch \cite{Balian76}, in the framework
of the theory of periodic orbits and classical quantization conditions
\cite{Gutz}.
In this approach the valence electrons
of the metal cluster are supposed to move in a spherical cavity
(representing the effects of the mean field). Supershells then occur from
the superposition of closed classical stationary orbits
\cite{Gutz} of the electrons,
the superposition of triangular and square orbits being the simplest
example \cite{Brec2271}.
The theory of Balian and Bloch leads to very specific predictions
about the behaviour of magic numbers, which can serve as a test for
alternative approaches. For example, the plot of $N_i^{1/3}$ vs. $i$,
where $N_i$ are the magic numbers of clusters of a specific alkali metal
and $i$ their index ($i=1$, 2, 3, \dots), should be a straight line
with a slope of 0.61 \cite{Brack677,Brec2271}.

In this paper we confront the results provided by the 3D $q$-HO to the
restrictions imposed by the theory of Balian and Bloch. It turns out that
agreement between the predictions of the two theories can be established
by introducing in the Hamiltonian of the 3D $q$-HO an additional term,
which is characterized by the same symmetry as the original Hamiltonian.
Furthermore, it can be seen that this additional term occurs naturally
through a variational procedure, analogous to the one used for obtaining
the Morse oscillator \cite{Morse} from the usual harmonic oscillator.

In Section 2 of this paper a brief account of the 3D $q$-HO is given,
while in Section 3 the application of the periodic orbit theory of Balian
and Bloch to metal clusters is briefly described. The predictions of
the 3D $q$-HO are compared to the restrictions imposed by the
theory of Balian and Bloch in Section 4, while in Section 5 a modified
Hamiltonian for the 3D $q$-HO is introduced, allowing for full
agreement with the theory of Balian and Bloch. Numerical details
concerning the modified 3D $q$-HO Hamiltonian are given in Section 6,
together with a study of supershells in the framework of this
Hamiltonian. In Section 7  a variational method leading from the usual
harmonic oscillator to the Morse oscillator is introduced, while
in Section 8 this method is applied for deriving the modified
Hamiltonian introduced in Section 5 from the original 3D $q$-HO
Hamiltonian. Finally in Section 9  a discussion of the present results
and plans for future work are given.

\section{The 3-dimensional $q$-deformed harmonic oscillator (3D $q$-HO)} %2

The space of the 3-dimensional $q$-deformed harmonic oscillator consists of
the completely symmetric irreducible representations of the quantum algebra
u$_q$(3) \cite{STK437,VdJL213,VdJ1799}.
In this space a deformed angular momentum algebra, so$_q$(3),
can be defined \cite{Terziev}.
The Hamiltonian of the 3D $q$-HO is defined so that it satisfies the following
requirements:

a) It is an so$_q$(3) scalar, i.e. the energy is simultaneously measurable
with the $q$-deformed  angular momentum related to the algebra so$_q$(3)
and its $z$-projection.

b) It conserves the number of bosons, in terms of which the quantum
algebras u$_q$(3) and so$_q$(3) are realized.

c) In the limit $q\to 1$ it is in agreement with the Hamiltonian of the usual
3-dimensional harmonic oscillator.

It has been proved \cite{Terziev} that a Hamiltonian of the 3D $q$-HO
satisfying the above requirements takes the form
\begin{equation} \label{eq:q1}
H_q = \hbar \omega_0 \left\{ [N] q^{N+1} - {q(q-q^{-1})\over [2] } C_q^{(2)}
\right\},
\end{equation}
where $N$ is the number operator and $C_q^{(2)}$ is the second order
Casimir operator of the algebra so$_q$(3), while
\begin{equation}\label{eq:q2}
[x]= {q^x-q^{-x} \over q-q^{-1}}
\end{equation}
is the definition of $q$-numbers and $q$-operators.
In the framework of quantum algebras in general, the deformation parameter 
$q$ can either be real ($q=e^\tau$, 
with $\tau$ being real) or a phase factor ($q=e^{i\tau}$, with $\tau$ 
being real). In the present case, however, the parameter $q$ is restricted 
to obtain real values only, since this is required in the derivation 
leading to the Hamiltonian of Eq. (1). (See the comment following Eq. (9) 
in Ref. \cite{JPA29} for more details.) 

The energy eigenvalues of the 3D $q$-HO are given by \cite{Terziev}
\begin{equation}\label{eq:q3}
E_q(n,l)= \hbar \omega_0 \left\{ [n] q^{n+1} - {q(q-q^{-1}) \over [2]}
[l] [l+1] \right\} = \hbar \omega_0 e_q(n,l),
\end{equation}
where $n$ is the number of vibrational quanta and $l$ is the eigenvalue of the
angular momentum, obtaining the values
$l=n, n-2, \ldots, 0$ or 1, and
\begin{equation}\label{eq:q4}
e_q(n,l)= [n] q^{n+1} - {q(q-q^{-1})\over [2]} [l] [l+1].
\end{equation}

In the limit of $q\to 1$ one obtains ${\rm lim}_{q\to 1} E_q(n,l)=
\hbar \omega_0 n$, which coincides with the classical result.

For small values of the deformation parameter $\tau$ (where $q=e^{\tau}$)
one can expand Eq. (3) in powers of $\tau$  obtaining \cite{Terziev}
$$
E_q(n,l)= \hbar \omega_0 n -\hbar \omega_0 \tau \left\{
l(l+1)-n(n+1)\right\}
$$
\begin{equation}\label{eq:q5}
-\hbar \omega_0 \tau^2 \left\{ l(l+1)-{1\over 3} n(n+1)(2n+1) \right\}
+ {\cal O} (\tau^3).
\end{equation}

The last expression to leading order bears great similarity to the modified
harmonic
oscillator suggested by Nilsson \cite{Nilsson1,Nilsson2}
(with the spin-orbit term omitted). Comparisons between the predictions
of the 3D $q$-HO and Nilsson's modified oscillator for the magic numbers
and supershells of metal clusters have been given in
\cite{PRA62,IJQC89,PRA65}. One of the main differences between the two models 
is that the term $\tau n(n+1)$ in the eigenvalues of the 3D $q$-HO given 
in Eq. (5) is occuring as a consequence of the overall symmetry, while 
the corresponding term $\mu' n(n+3)/2$ in the eigenvalues of Nilsson's 
modified oscillator (see, for example, Eq. (3) of Ref. \cite{PRA65}) is put 
in ``by hand''. It should also be noticed that the use of $q$-deformations 
does not mean that an additional parameter is introduced in the theory.
In both the 3D $q$-HO and Nilsson's modified oscillator only one parameter 
appears (in addition to the overall scale).  
In the 3D $q$-HO case this is the deformation parameter $\tau$, while 
in Nilsson's modified oscillator it is the parameter $\mu'$ 
(see, for example, Eq. (3) of Ref. \cite{PRA65}). 

\section{The theory of Balian and Bloch} %3

The theory of Balian and Bloch \cite{Balian76}
is an example of periodic orbit theory \cite{Gutz},
developed as a semiclassical bridge between quantum mechanics and
classical mechanics.
In the approach of Balian and Bloch, shell effects in metal clusters
are studied by considering the valence electrons moving on straight
lines within a smooth sphere and being reflected on the inner surface
of the sphere. Periodic orbits in this case correspond to various polygons
with three or more corners, the triangular and the square orbits being the
simplest  examples \cite{Brack677,Brec2271}.

In the theory of Balian and Bloch a cluster of $N$ alkali atoms can be studied
\cite{Brec2271} by considering the $N$ valence electrons moving in a sphere
with radius
\begin{equation}\label{eq:q6}
R=r_S N^{1/3},
\end{equation}
where $r_S$ is the Wigner--Seitz radius. The lengths of the triangular
and the square orbits are then respectively
\begin{equation} \label{eq:q7}
L_3  = 3 \sqrt{3} r_S N^{1/3},
\end{equation}
\begin{equation} \label{eq:q8}
L_4 = 4 \sqrt{2} r_S N^{1/3}.
\end{equation}
Assuming that electrons move with the Fermi velocity $v_F$,
the beating pattern created by the superposition of the two orbits leads
to supershell structure. It turns out that the magic numbers (shell
closures) $N_i$ are related to the index $i$ counting their number
($i=1$, 2, 3, \dots) by \cite{Brec2271}
\begin{equation} \label{eq:q9}
N_i^{1/3} = {h \over m v_F r_S} \quad {2\over 3 \sqrt{3}+ 4 \sqrt{2} }
\quad  i =0.605 i,
\end{equation}
where $m$ is the electron mass and the fact that $v_F r_S$ is constant
for all metals has been used. Therefore the plot of $N_i^{1/3}$ vs. $i$
should be a straight line with a slope of 0.61~.
Furthermore,  a phase shift of  a half unit of the index $i$
should occur in the node region of the beat pattern, i.e., when passing
from a supershell to the next one \cite{Brack677,Brec2271}.
In addition, within each supershell the shell closures should appear
at equidistant positions (i.e., they should exhibit a periodicity) when
plotted vs. $N^{1/3}$ \cite{Brack677}, where $N$ is the number of
particles (valence electrons in the present case).

Comparisons to experimental sets of magic numbers have shown
\cite{Brack677} that a slope of 0.61 is obtained in the cases of Na and Li
clusters, revealing that the triangular and square orbits represent a good
approximation in these cases.
For Al clusters, though, a slope of 0.32 has been obtained \cite{Brack677},
indicating that
in this case, if the theory is applicable, more complicated orbits enter
\cite{Lerme2818}.

\section{Comparing the predictions of the 3D $q$-HO and of the Balian--Bloch
theory} %4

As we have seen in the previous section, a hallmark of the theory of
Balian and Bloch is that the plot of $N_i^{1/3}$ vs. $i$, where
$N_i$ are the magic numbers and $i$ the index counting them
($i=1$, 2, 3, \dots), should be a straight line having a slope of
0.61 in the case of alkali metals, while in the case of Al clusters
the slope should be 0.32 \cite{Brack677}.

In order to compare the predictions of the 3D $q$-HO to the restrictions
imposed by the theory of Balian and Bloch, we plot in Fig. 1(a)
(line labelled by $\epsilon=0.0$) the magic
numbers obtained from the 3D $q$-HO for $\tau=0.038$, the parameter value
found appropriate in Ref. \cite{PRA62} for reproducing the magic numbers of
alkali clusters (up to 1500 atoms, which is the limit of validity of theories
based on the filling of electronic shells \cite{Martin209,Clem44}).
Numbers considered as magic, listed in Table 1, correspond to energy gaps
larger than $\delta=0.38$, as in Ref. \cite{PRA62}, with
$\hbar \omega_0=1$.
In the same figure, a straight line with a slope of 0.61
appears. It is clear that the 3D $q$-HO magic numbers follow the straight
line up to $i=14$ quite well, while beyond this point the predictions
of the 3D $q$-HO are clearly lower than the straight line, indicating
that ``too many'' magic numbers are produced by the model in this region.

The same conclusion is arrived at by looking at Fig. 1(b), where
the predictions of the 3D $q$-HO for $\tau=0.050$, the parameter value
found in Ref. \cite{PRA62} appropriate for reproducing the magic numbers
of Al clusters, are reported (line labelled by $\epsilon=0.0$).
Again magic numbers, listed in Table 1,
correspond to energy gaps larger than $\delta =0.38$ (with $\hbar
\omega_0=1$). Small magic numbers (below
186, i.e., below $i=9$) are not shown, since it is known that small
magic numbers in Al clusters cannot be explained by models based on the
filling of electronic shells, because of the symmetry breaking caused by
the ionic lattice \cite{Persson},
while for large magic numbers this problem does not exist.
In the plot a straight line with a slope of 0.32, which is expected
to be appropriate for Al clusters \cite{Brack677},
as mentioned in the previous section,
is also seen. It is clear that the predictions of the 3D $q$-HO follow
a line parallel to the one with slope 0.32 roughly up to $i=23$, while
beyond this point the slope is gradually reduced, indicating that
the model predicts ``too many'' magic numbers.

\section{A modified Hamiltonian for the 3D $q$-HO} %5

The discrepancy between the 3D $q$-HO and the theory of Balian
and Bloch can be lifted by considering the Hamiltonian
\begin{equation} \label{eq:q10}
 H'_q = H_q -\epsilon H_q^2,
\end{equation}
with eigenvalues
\begin{equation} \label{eq:q11}
E'_q(n,l)= E_q(n,l)-\epsilon E_q^2 (n,l),
\end{equation}
where $\epsilon$ is a small real positive constant.
Justification for this choice will be given in  Sections 7 and 8
through a variational procedure. For the moment the following comments
suffice:

a) It is clear that $H'_q$ is a function of $H_q$, which is by construction
an so$_q$(3) scalar, as mentioned in Section 2. Therefore $H'_q$ is also
an so$_q$(3) scalar.

b) The energy eigenvalues of the new Hamiltonian can be written in the form
\begin{equation} \label {eq:q12}
E'_q(n,l) = \hbar \omega_0 \left( 1-\epsilon \hbar \omega_0 e_q(n,l)
\right) e_q(n,l)  = \hbar \omega(n,l) e_q(n,l),
\end{equation}
where
\begin{equation}\label{eq:q13}
\omega(n,l)= \omega_0 \left(1-\epsilon \hbar \omega_0 e_q(n,l)\right)
\end{equation}
is a variable frequency, depending on the quantum numbers $n$, $l$,
and on the small parameter $\epsilon$. We shall call this oscillator
the {\sl Variable Frequency Oscillator} (VFO) corresponding to
the 3D $q$-HO, a term for which
justification will be provided in Sections 7 and 8.

The magic numbers provided by the VFO for a few appropriate values
of $\epsilon$ in the case of $\tau=0.038$, which is relevant for alkali
clusters \cite{PRA62}, are shown in Fig. 1(a) and listed in Table 1.
Details of the calculation will be given in Section 6.
Once more magic numbers are separated
by gaps larger than $\delta=0.38$, while $\hbar\omega_0=1$.
It is clear that the predictions of the VFO roughly follow the straight
line with a slope of 0.61 even for large values of $i$, thus overcoming
the difficulties faced by the 3D $q$-HO.

A similar picture is obtained for $\tau=0.050$ (and $\delta=0.38$, with
$\hbar \omega_0=1$), which is appropriate for Al clusters \cite{PRA62}.
For a few appropriate values of $\epsilon$, shown in Fig. 1(b) and listed
in Table 1,  the predictions of the VFO roughly follow a straight line
with a slope of 0.32, as they should, according to the previous section.

We therefore conclude that the addition of the second term in the
Hamiltonian of the 3D $q$-HO, leading to the VFO, makes the predictions
of the 3D $q$-HO  compatible with the predictions of the theory of Balian
and Bloch.

\section{Numerical details} %6

In this section the calculations leading to the results reported
in Section 5 will be described.
Throughout this paper we put $\hbar \omega_0 =1$ for simplicity.

An important difference between the 3D $q$-HO and the VFO of Eq. (\ref{eq:q10})
lies in the way truncations of the spectrum are made. The following
comments apply:

a) In the 3D $q$-HO the level with $l=n$ always lies lowest in energy within
each shell, the level with $l=n-2$ lies immediately above it, and so on.
Therefore stopping the level scheme at the $l=n$ level of a given
shell and taking  into account all levels with lower $n$ (i.e., all levels
of the shells lying below the given one), one makes sure that all levels
up to the given level have been included \cite{PRA65}.

b) In the case of the VFO of Eq. (\ref{eq:q10}) the following picture occurs:
For a given (small) value of $\epsilon$ the first several shells
exhibit the same behaviour as in the case of the 3D $q$-HO, i.e., the level
with $l=n$ lies lowest in energy within each shell, the level with
$l=n-2$ lies immediately above it, and so on. As the shell number
is increasing, however, an inversion of the order of the levels occurs,
with the levels with $l=0$ (for $n$ being even) or $l=1$ (for $n$ being odd)
lying lowest in energy within the shell. It is then clear that beyond this
inversion point truncation at a given shell should be made at the level with
$l=0$ (for $n$ being even) or at $l=1$ (for $n$ being odd).

The magic numbers obtained for the values of $\tau$ and $\epsilon$ used
in the previous section are given in Table 1. In each case the maximum value
of $n$, $n_{max}$, included in the calculation is indicated. Care has been
taken that no inversion of the order of the levels, of the type described
in comment b) above, occurs for the values
of $n$ included in the calculation. Therefore in all cases truncation is made
at the level with $n=n_{max}$ and $l=n_{max}$. The total number of levels up
to the truncation point, $N_{max}$, is also shown in Table 1. The following
remarks are now in place:

a) In the cases considered here and up to the truncation point, for a given
value of $\tau$ the order of the levels is not modified as $\epsilon$ is
changed. The only modification occuring is that the spectrum gets ``squeezed''
as $\epsilon$ increases.

b) As a result of a), the magic numbers reported in Table 1 present the
following feature. For each value of $\tau$ and for $\epsilon =0.0$
the largest number of magic numbers appears. As $\epsilon$ increases,
some of the magic numbers cease to be magic any more, since the
``squeezing'' of the spectrum brings the levels closer to each other.
For each value of $\epsilon$ the magic numbers occuring are a subset
of the magic numbers occuring for lower values of $\epsilon$ (with
the same $\tau$). No new magic numbers appear, within the limits
considered here, as $\epsilon$ increases.

It is interesting to examine at this point what the influence
of the additional term to the appearance of supershells is.
For this purpose we are going to use the procedure employed by Nishioka
{\it et al.} \cite{Nishi,Nishi2}.
For a given number of particles $N$ the single particle
energies $E_j(n,l)$ of the $N$ occupied states are summed up
\begin{equation}\label{eq:q14}
E(N)= \sum_{j=1}^N E_j(n,l).
\end{equation}
This sum is then divided into two parts: A smooth average part $E_{av}$
and a shell part $E_{shell}$, which will exhibit the supeshell structure
\begin{equation}\label{eq:q15}
E(N)= E_{av}(N)+E_{shell}(N).
\end{equation}
For the average part of the total energy a Liquid Drop Model expansion
is used \cite{PRA65}
\begin{equation} \label{eq:q16}
E_{av}(N)= a_1 N^{1/3} + a_2 N^{2/3} + a_3 N + a_4 N^{4/3}
+ a_5 N^{5/3} + a_6 N^2.
\end{equation}

The parameters of the fits occuring in the cases considered here
are shown in Table 2, together with the number of levels, $N_{max}$,
included in the fit and the rms deviation $\sigma$. We remark that
for a given value of $\tau$ the parameters change smoothly for the
nonzero values of $\epsilon$, while the case with $\epsilon=0.0$
is characterized by quite different values of the parameters, but also
by a higher rms deviation $\sigma$. The addition of the second term
in Eq. (\ref{eq:q10}) improves the agreement of the average part of the total
energy to the Liquid Drop Model expansion, thus resulting in
lower rms deviations $\sigma$.

The procedure of the calculation was as follows: First the summations
described by Eq. (\ref{eq:q14}), resulting in the total energy $E(N)$ for each
particle number $N$, have been performed. Subsequently, in order to reduce
the size of the calculation approximately by a factor of 10, the average
$E(N)$ was calculated every 11 points (i.e., for $N=6$, 17, 28, \dots)
up to the cutoff point which is reported
in Table 2 as $N_{max}$. These averaged values of $E(N)$ were
subsequently fitted by the expansion of Eq. (\ref{eq:q16}), resulting in the
determination of $E_{av}(N)$ at these points. Finally $E_{shell}(N)$ has
been obtained at these points as the difference $E(N)-E_{av}(N)$
and plotted in Figs. 2 and 3.

The shell energy, $E_{shell}$, is plotted vs. the particle
number, $N$,  for $\tau=0.038$, which is appropriate for alkali clusters
\cite{PRA62}, in Fig. 2. In Fig. 2(a) the predictions of the
original 3D $q$-HO (with $\epsilon=0$) are shown, while in Fig. 2(b)
the results of the VFO with $\epsilon=0.006$ are depicted. The VFO with
$\epsilon=0.007$, 0.008 gives results which look almost identical
with Fig. 2(b) and therefore are not shown for brevity.
The similarities between Figs. 2(a) and 2(b) are clear. Not only the
supershell appears in both cases around $N=1000$, as it is expected
for Na clusters \cite{Nishi,Nishi2},
but in addition even the maxima and minima of the shell
energy appear at the same particle numbers and have roughly the same
magnitude. Even the local maxima and minima present striking similarities.
These results corroborate the remarks made above, namely that the addition
of the second term to the Hamiltonian of Eq. (\ref{eq:q10})
does not influence the order of the energy levels,
the main effect of the second term being the gradual ``squeezing'' of the
spectrum as energy increases. Of course this conclusion is valid only
within the region of particle numbers studied and for small values
of $\epsilon$, like the ones used here.

Similar results are obtained in Fig. 3 for the case of $\tau=0.050$,
which is appropriate for Al clusters \cite{PRA62}.
Besides the $\epsilon =0$ case,
shown in Fig. 3(a),
the results corresponding to $\epsilon=0.0050$ are shown in Fig. 3(b),
since the
cases with $\epsilon=0.0053$, 0.0055 provide results almost identical
with the ones shown in Fig. 3(b). In all cases there is some evidence
for a supershell below $N=1000$, although its appearance is not as clear
as in the case of Fig. 2. The appearance of a supershell in this region
is in agreement with the results of more sophisticated calculations,
as, for example, spherical jellium model predictions in  Local Density
Approximation \cite{Genzken}, but it is not in good agreement with
experiment, where no evidence for supershell in Al clusters exists
in this region \cite{Pell96,Pell98}. An advantage of the VFO
in comparison to earlier calculations \cite{Genzken}
is that at least it can reproduce
the slope of 0.32 in Fig. 1(b), something which is not occuring
in spherical jellium model calculations, although it occurs
experimentally \cite{Brack677,Pell96,Pell98}.

In the theory of Balian and Bloch, as mentioned above, a phase shift by
a half unit of the running index $i$ should be observed in the plot
of the magic numbers $N_i^{1/3}$ vs. $i$ when passing from a supershell
to the next \cite{Brack677,Brec2271}.
In Fig. 1(a) such a shift is seen quite clearly around
$N_i=1000$ in the cases of $\epsilon=0.006$, 0.007, 0.008, while
no clear shift of this type is seen in Fig. 1(b).

Furthermore, the theory of Balian and Bloch, as mentioned above,
predicts that within each supershell the minima of the shell energy,
$E_{shell}$, should appear at equidistant positions (i.e. they
should exhibit a periodicity) when plotted vs. $N^{1/3}$. The change
in the periodicity when passing from the first supershell to the second one
is clear in Figs. 2 and 3 (although in these cases, for reasons of clarity,
$E_{shell}$ is plotted vs. $N$ and not vs. $N^{1/3}$). The fact that the
predictions of the VFO corresponding to the 3D $q$-HO approximately exhibit
the right periodicity features does not come as a surprise, since
the 3D $q$-HO is known to show this feature \cite{IJQC89,PRA65},
while, as we have seen above, the addition of the second term in Eq.
(\ref{eq:q10}) does not influence the position of the minima.

In conclusion, the VFO corresponding to the 3D $q$-HO is able to
reproduce the right slope in the $N_i^{1/3}$ vs. $i$ plot in both
the alkali and Al clusters. In addition it predicts correctly
the first supershell in alkali clusters, while in Al clusters
its prediction for a supershell is in rough agreement with
results of spherical jellium models but not with experiment.

\section{A variational method} %7

In nuclear physics it is well known that nuclear spectra can be described
very accurately in terms of the Variable Moment of Inertia (VMI) model
\cite{VMI}. In this model the energy levels are given by
\begin{equation} \label{eq:q17}
E(J)= {J(J+1)\over 2\Theta(J)} +{1\over 2} C \{\Theta(J)-\Theta(0)\}^2,
\end{equation}
where $J$ is the angular momentum and $\Theta(J)$ is the moment of inertia,
which is supposed to be a function of the angular momentum. $C$ and
$\Theta(0)$ are free parameters, the latter representing the ground state
moment of inertia. It is clear that the VMI formula is a generalization
of the rigid rotator formula
\begin{equation} \label{eq:q18}
E(J)= {J(J+1)\over 2\Theta},
\end{equation}
in which the moment of inertia is assumed to be constant. The rigid rotator
formula is known to fail beyond the first few levels of a rotational nucleus,
since the experimental levels appear ``squeezed'' in comparison to the
rigid rotator predictions. This difficulty is overcome in the framework
of the VMI model by determining the moment of inertia for each
value of the angular momentum $J$ through a minimization of the energy
with respect to the moment of inertia for given angular momentum
\begin{equation} \label{eq:q19}
{\partial E(J) \over \partial \Theta(J)} |_J =0.
\end{equation}
This variational condition leads to a cubic equation for $\Theta(J)$,
which turns out to have only one real solution \cite{VMI}, corresponding
to the
appropriate value of the moment of inertia for the given value of the
angular momentum.  The second term in Eq. (\ref{eq:q17})
is justified by the well
known fact that many perturbing  potentials near their origin can be
approximated by a harmonic oscillator potential.

Following the same reasoning, it is interesting to examine what happens
to the usual harmonic oscillator if, by analogy, one allows the angular
frequency to be a function of the quantum number $n$. The energy
will then read
\begin{equation} \label{eq:q20}
E(n)= \hbar \omega(n) \left( n+{1\over 2}\right) + {1\over 2} C
\{\omega(n)-\omega(0)\}^2,
\end{equation}
where $C$ and $\omega(0)$ are free parameters, the latter corresponding
to the ground state angular frequency. The variational principle in this
case should correspond to the minimization of the energy with respect to the
angular frequency for constant value of the quantum number $n$
\begin{equation} \label{eq:q21}
{\partial E(n) \over \partial \omega(n) } |_n =0.
\end{equation}
It is clear that this condition leads to
\begin{equation} \label{eq:q22}
\hbar \left( n+{1\over 2}\right) + C \{\omega(n)-\omega(0)\}=0
\Rightarrow \omega(n)=\omega(0)-{\hbar \over C} \left(n+{1\over 2}\right) .
\end{equation}
Substituting this result in Eq. (\ref{eq:q20}) we obtain
\begin{equation} \label{eq:q23}
E(n) = \hbar \omega(0) \left( n+{1\over 2}\right) -{1\over 2} {\hbar^2
\over C} \left( n+{1\over 2}\right)^2,
\end{equation}
which is reminiscent of the spectrum of the Morse potential \cite{Morse}.

Indeed, solving the Schr\"odinger equation for the Morse potential
\cite{Flugge,Cooper112,Cooper25}
\begin{equation}\label{eq:q24}
V(x)= D (1 -e^{-\alpha x})^2,
\end{equation}
one obtains the energy spectrum
\begin{equation} \label{eq:q25}
E(n) = \hbar \omega \left\{ \left( n+{1\over 2}\right) -x_e
\left( n+{1\over 2}\right)^2\right\},
\end{equation}
where
\begin{equation} \label{eq:q26}
 x_e = {1\over 2} {\hbar \alpha \over \sqrt{2 m D}},
\end{equation}
and
\begin{equation} \label{eq:q27}
\omega =\alpha \sqrt{ 2D \over m}.
\end{equation}

We therefore conclude that by allowing the angular frequency of the simple
harmonic oscillator to vary with the quantum number $n$, we obtain the
spectrum of the Morse oscillator. Again the second term in Eq. (\ref{eq:q20})
is
in agreement to the fact that most perturbing potentials near their origin
can be approximated by the harmonic oscillator potential. We shall refer
to the oscillator of Eq. (\ref{eq:q20})
as the {\sl Variable Frequency Oscillator} (VFO).

\section{Derivation of the modified Hamiltonian for the 3D $q$-HO \\
through a variational method} %8

The idea leading to the VFO of the previous section can be appropriately
generalized in the case of the 3D $q$-HO. In this case we consider
the energy expression
$$
E'_q(n,l)= \hbar \omega(n,l) \left\{ [n]q^{n+1}-{ q(q-q^{-1})\over [2]}
[l][l+1]\right\} + {1\over 2} C \{\omega(n,l)-\omega(0,0)\}^2
$$
\begin{equation} \label{eq:q28}
= \hbar \omega(n,l) e_q(n,l) +{1\over 2} C \{\omega(n,l)-\omega(0,0)\}^2,
\end{equation}
where the angular frequency $\omega(n,l)$ depends on the quantum
numbers $n$ and $l$, while $C$ and $\omega(0,0)$ are real positive
constants, the latter corresponding to the ground state angular
frequency, since the ground state of the 3D $q$-HO is characterized by
$n=0$ and $l=0$. It is thus clear that $\omega(0,0)$ corresponds to
$\omega_0$ appearing in Eq. (\ref{eq:q3}), i.e., $\omega(0,0)\equiv
\omega_0$.
The variational condition in the present case
should read
\begin{equation} \label{eq:q29}
 {\partial E'_q(n,l) \over \partial \omega(n,l)} |_{n,l} =0.
\end{equation}
In other words, the energy is minimized with respect to the angular
frequency for constant values of the quantum numbers $n$ and $l$.
The variational condition leads to the equation
$$
\hbar e_q(n,l) +C \{\omega(n,l)-\omega(0,0)\}=0 \Rightarrow
\omega(n,l)= \omega(0,0)-{\hbar \over C} e_q(n,l)
$$
\begin{equation}\label{eq:q30}
 =\omega(0,0)
-{\hbar \over C} \left\{ [n] q^{n+1} - {q(q-q^{-1}) \over [2]}
[l][l+1]\right\}.
\end{equation}
Substituting this result in Eq. (\ref{eq:q28}) one then obtains
\begin{equation} \label{eq:q31}
E'_q(n,l)= \hbar \omega(0,0) e_q(n,l) - {1\over 2} {\hbar^2 \over C}
e^2_q(n,l),
\end{equation}
which is the same as Eq. (\ref{eq:q11}), with
\begin{equation} \label{eq:q32}
\epsilon= {1\over 2 C \omega_0^2},
\end{equation}
since $\omega_0 \equiv \omega(0,0)$, as mentioned above.

\section{Discussion} %9

In this paper we have attempted a comparison of the predictions
for the shell structure of metal clusters
of the 3D $q$-HO model to the ones of the periodic orbit theory
of Balian and Bloch. It turns out that the predictions of the
3D $q$-HO for the magic numbers of metal clusters can be made
compatible with the predictions of the theory of Balian and Bloch
by adding to the 3D $q$-HO a symmetry-preserving correction term reminiscent
of the anharmonicity term in the spectrum of the Morse potential, while
this addition does not influence the predictions for the supershells.
This extended expression for the 3D $q$-HO can be
justified through a variational method, similar to the one used in the
Variable Moment of Inertia (VMI) model of nuclear physics, leading to the
concept of the Variable Frequency Oscillator (VFO), which gives promise
of wider applicability.

\bigskip
\noindent
{\bf Acknowledgements}
\medskip

One of the authors (PPR) acknowledges support from the Bulgarian Ministry
of Science and Education under contracts $\Phi$-415 and $\Phi$-547.

\bs\bs 

\centerline{\bf Figure captions} 

{\bf Fig. 1}  Cubic roots of the magic numbers $N_i$ plotted vs. the
running index $i$ counting them. The magic numbers are the ones listed
in Table 1. (a) $\tau=0.038$, (b) $\tau=0.050$.

{\bf Fig. 2} Shell part ($E_{shell}$) of the total energy [in units of
$\hbar \omega_0$, see Eqs. (\ref{eq:q3}) and (\ref{eq:q11})]
vs. the number of particles $N$, in the case of $\tau=0.038$,
(a) for the 3D $q$-HO, (b) for a corresponding VFO. The values of the
dimensionless parameters $\tau$ and $\epsilon$ are listed in Table 2,
together with the details of the calculation. See Section 6
for further discussion.

{\bf Fig. 3} Same as Fig. 2, but for $\tau=0.050$.

\newpage 

%%%%%%%%%%%%%%%%%%%%%%%%%%%%%%%%%%%%%%%%%%%%%%%%%%%%%%%%%%%%%%%%%%%%%%
%%%%%%%%%%%%%%%%%%% Table 1 %%%%%%%%%%%%%%%%%%%%%%%%%%%%%%%%%%%%%%%%

\begin{table}

\begin{center}

\caption{Magic numbers (corresponding to gaps larger than $\delta=0.38$,
with $\hbar \omega_0=1$)
produced by the 3D $q$-HO [Eq. (\ref{eq:q3})]
(cases with $\epsilon=0.0$) and the
corresponding VFO [Eq. (\ref{eq:q11})] for different values of the parameters
$\tau$ and $\epsilon$. $n_{max}$ is the maximum value of the quantum
number $n$ included in the calculation, while the highest level
taken into account is the one with $n=n_{max}$ and $l=n_{max}$,
corresponding to the reported total number of particles $N_{max}$.
$i$ is a running index counting the magic numbers. See Section 6
for further discussion. }

\begin{tabular}{r r r r r r r r r }
\hline
$\tau$   &0.038 & 0.038 & 0.038 & 0.038 & 0.050 & 0.050  & 0.050  & 0.050  \\
$\epsilon$& 0.0 & 0.006 & 0.007 & 0.008 & 0.0   & 0.0050 & 0.0053 & 0.0055 \\
$n_{max}$ & 26  &  26   & 26    & 25    & 26    &  26    &  25    &  25    \\
$N_{max}$ &4658 & 4658  & 4658  & 4154  & 4778  & 4778   & 4258   & 4258   \\
\hline
$i$&     &     &     &     &     &     &     &     \\
 1 &   2 &   2 &   2 &   2 &   2 &   2 &   2 &   2 \\
 2 &   8 &   8 &   8 &   8 &   8 &   8 &   8 &   8 \\
 3 &  20 &  20 &  20 &  20 &  20 &  20 &  20 &  20 \\
 4 &  34 &  34 &  40 &  40 &  34 &  34 &  34 &  34 \\
 5 &  40 &  40 &  58 &  58 &  40 &  40 &  40 &  40 \\
 6 &  58 &  58 &  92 &  92 &  58 &  58 &  58 &  58 \\
 7 &  92 &  92 & 138 & 138 &  92 &  92 &  92 &  92 \\
 8 & 138 & 138 & 198 & 198 & 138 & 138 & 138 & 138 \\
 9 & 198 & 198 & 254 & 254 & 186 & 186 & 186 & 186 \\
10 & 254 & 254 & 338 & 338 & 254 & 254 & 254 & 254 \\
11 & 268 & 338 & 440 & 440 & 338 & 338 & 338 & 338 \\
12 & 338 & 440 & 676 & 676 & 398 & 398 & 398 & 398 \\
13 & 440 & 676 & 832 & 832 & 440 & 440 & 440 & 440 \\
14 & 556 & 832 & 912 & 912 & 486 & 542 & 542 & 542 \\
15 & 562 & 912 &1012 &1012 & 542 & 612 & 612 & 612 \\
16 & 676 &1012 &1100 &1100 & 612 & 676 & 676 & 676 \\
17 & 694 &1100 &1206 &1206 & 676 & 748 & 748 & 748 \\
18 & 832 &1206 &1660 &1660 & 748 & 832 & 832 & 832 \\
19 & 912 &1660 &1760 &1760 & 832 & 912 & 912 & 912 \\
20 &1012 &1760 &2048 &2048 & 890 &1006 &1006 &1006 \\
21 &1100 &2048 &2368 &2368 & 912 &1074 &1074 &1074 \\
22 &1206 &2368 &3028 &3028 &1006 &1100 &1100 &1100 \\
23 &1284 &3028 &3438 &3438 &1074 &1284 &1284 &1284 \\
24 &1314 &3438 &3886 &3886 &1100 &1314 &1314 &1410 \\
25 &1410 &3886 &4374 &     &1206 &1410 &1410 &1502 \\
26 &1502 &4052 &     &     &1284 &1502 &1502 &1760 \\
27 &1516 &4374 &     &     &1314 &1516 &1760 &2018 \\
28 &1660 &     &     &     &1410 &1760 &2018 &2048 \\
29 &1760 &     &     &     &1502 &2018 &2048 &2178 \\
30 &2018 &     &     &     &1516 &2048 &2178 &2334 \\
\hline
\end{tabular}
\end{center}
\end{table}

\newpage 

%%%%%%%%%%%%%%%%%%%%%%%%%%%%%%%%%%%%%%%%%%%%%%%%%%%%%%%%%%%%%%%%%%%%%%
%%%%%%%%%%%%% Table 1 (continued) %%%%%%%%%%%%%%%%%%%%%%%%%%%%%%%%%%%%%

\setcounter{table}{0}

\begin{table}

\begin{center}

\caption{ (continued)}

\begin{tabular}{r r r r r r r r r }
\hline
$\tau$   &0.038 & 0.038 & 0.038 & 0.038 & 0.050 & 0.050  & 0.050  & 0.050  \\
$\epsilon$& 0.0 & 0.006 & 0.007 & 0.008 & 0.0   & 0.0050 & 0.0053 & 0.0055 \\
$n_{max}$ & 26  &  26   & 26    & 25    & 26    &  26    &  25    &  25    \\
$N_{max}$ &4658 & 4658  & 4658  & 4154  & 4778  & 4778   & 4258   & 4258   \\
\hline
$i$&     &     &     &     &     &     &     &     \\
31 &2048 &     &     &     &1614 &2178 &2334 &2368 \\
32 &2178 &     &     &     &1660 &2334 &2368 &2510 \\
33 &2334 &     &     &     &1734 &2368 &2510 &2672 \\
34 &2368 &     &     &     &1760 &2510 &2672 &2722 \\
35 &2654 &     &     &     &1778 &2672 &2722 &3028 \\
36 &2672 &     &     &     &1940 &2722 &3028 &3050 \\
37 &2722 &     &     &     &2018 &3028 &3050 &3112 \\
38 &2796 &     &     &     &2048 &3050 &3112 &3438 \\
39 &3028 &     &     &     &2178 &3112 &3438 &3464 \\
40 &3050 &     &     &     &2334 &3438 &3464 &3886 \\
41 &3190 &     &     &     &2368 &3464 &3886 &3916 \\
42 &3404 &     &     &     &2510 &3886 &3916 &     \\
43 &3438 &     &     &     &2672 &3916 &     &     \\
44 &3464 &     &     &     &2684 &3988 &     &     \\
45 &3610 &     &     &     &2722 &4374 &     &     \\
46 &3848 &     &     &     &2876 &4408 &     &     \\
47 &3886 &     &     &     &3028 &     &     &     \\
48 &4052 &     &     &     &3050 &     &     &     \\
49 &4312 &     &     &     &3112 &     &     &     \\
50 &4326 &     &     &     &3190 &     &     &     \\
51 &4374 &     &     &     &3244 &     &     &     \\
52 &4552 &     &     &     &3438 &     &     &     \\
53 &     &     &     &     &3464 &     &     &     \\
54 &     &     &     &     &3528 &     &     &     \\
55 &     &     &     &     &3622 &     &     &     \\
56 &     &     &     &     &3680 &     &     &     \\
57 &     &     &     &     &3886 &     &     &     \\
58 &     &     &     &     &3916 &     &     &     \\
59 &     &     &     &     &3988 &     &     &     \\
60 &     &     &     &     &4088 &     &     &     \\
61 &     &     &     &     &4156 &     &     &     \\
62 &     &     &     &     &4374 &     &     &     \\
63 &     &     &     &     &4408 &     &     &     \\
64 &     &     &     &     &4462 &     &     &     \\
65 &     &     &     &     &4488 &     &     &     \\
66 &     &     &     &     &4578 &     &     &     \\
67 &     &     &     &     &4596 &     &     &     \\
\hline
\end{tabular}
\end{center}
\end{table}

\newpage 

%%%%%%%%%%%%%%%%%%%%%%%%%%%%%%%%%%%%%%%%%%%%%%%%%%%%%%%%%%%%%%%%%%%%%%
%%%%%%%%%%%%%%%%%%% Table 2 %%%%%%%%%%%%%%%%%%%%%%%%%%%%%%%%%%%%%%%%

\begin{table}

\begin{center}

\caption{
Parameters used for fitting the average part of the total
energy [see Eq. (\ref{eq:q16})] in the case of the 3D $q$-HO and the
corresponding VFO
for various values of the parameters $\tau$ and $\epsilon$,
corresponding to the cases exhibited in Figs. 2 and 3.
The parameters are dimensionless, since we have assumed $\hbar \omega_0=1$
[see Eqs. (\ref{eq:q3}) and (\ref{eq:q11})] throughout. The number of
particles
$N_{max}$ included in each calculation and the relevant rms deviation
$\sigma$ are also shown. See Section 6 for further discussion.
}

\begin{tabular}{r r r r r r r r r r }
\hline
$ 10^3 \tau$&$ 10^4 \epsilon$&
$a_1$  & $a_2$& $a_3$& $10 a_4$& $10^2 a_5$& $10^4 a_6$ & $N_{Max}$
& $\sigma$ \\
\hline
38&  0&-21.035& 18.295& -7.295&19.521& -6.082& 40.857&3009&8.904\\
38& 60& 24.756&-20.883&  5.201& 0.493&  7.946& -8.993&3009&5.758\\
38& 70& 32.475&-27.496&  7.306&-2.704& 10.297&-17.326&3009&5.297\\
38& 80& 39.762&-33.786&  9.329&-5.806& 12.597&-25.559&3009&4.834\\
50&  0&-24.946& 24.641&-10.384&26.117&-13.000& 82.558&2008&7.328\\
50& 50& 14.051&-13.208&  3.323& 2.409&  7.006&  6.484&2008&5.286\\
50& 53& 16.795&-15.857&  4.264& 0.817&  8.320&  1.634&2008&5.175\\
50& 55& 18.589&-17.556&  4.862&-0.188&  9.150& -1.464&2008&5.098\\
\hline
\end{tabular}
\end{center}
\end{table}


\begin{thebibliography}{99}

\bibitem{Iac1}
F. Iachello and A. Arima, {\it The Interacting Boson Model} (Cambridge
University Press, Cambridge, 1987).

\bibitem{Iac2}
F. Iachello and R. D. Levine, {\it Algebraic Theory of Molecules}
(Oxford University Press, Oxford, 1995).

\bibitem{deHeer611}
W. A. de Heer, {\it Rev. Mod. Phys.} {\bf 65}, 611-676 (1993).

\bibitem{Brack677}
M. Brack, {\it Rev. Mod. Phys.} {\bf 65}, 677-732 (1993).

\bibitem{PRA62}
D. Bonatsos, N. Karoussos, D. Lenis, P. P. Raychev, R. P. Roussev and
P. A. Terziev, {\it Phys. Rev. A} {\bf  62}, 013203 [13 pages] (2000).

\bibitem{IJQC89}
D. Bonatsos, D. Lenis, P. P. Raychev and P. A. Terziev, {\it Int. J. Quant.
Chem.} {\bf 89}, 299-312 (2002).

\bibitem{PRA65}
D. Bonatsos, D. Lenis, P. P. Raychev and P. A. Terziev, {\it Phys. Rev. A}
{\bf 65}, 033203 [12 pages] (2002).

\bibitem{AQC40}
A. I. Kuleff, J. Maruani and P. P. Raychev,  {\it Adv. Quant. Chem.} {\bf 40},
279-304 (2001).

\bibitem{IJQC377}
D. Bonatsos, A. I. Kuleff, J. Maruani, P. P. Raychev and P. A. Terziev,
{\it Int. J. Quant. Chem.} {\bf 89}, 377-388 (2002).

\bibitem{Terziev}
P. P. Raychev, R. P. Roussev, N. Lo Iudice and P. A. Terziev, {\it J. Phys. G:
Nucl. Part. Phys.} {\bf 24}, 1931-1943 (1998).

\bibitem{PPNP}
D. Bonatsos and C. Daskaloyannis, {\it Prog. Part. Nucl. Phys.} {\bf 43},
537-618 (1999).

\bibitem{STK437}
Yu. F. Smirnov, V. N. Tolstoy and Yu. I. Kharitonov, {\it Yad. Fiz.} {\bf 54},
721-736 (1991) [{\it Sov. J. Nucl. Phys.} {\bf 54}, 437-445 (1991)].

\bibitem{VdJL213}
J. Van der Jeugt, {\it J. Phys. A: Math. Gen.} {\bf 25}, L213-L218 (1992).

\bibitem{VdJ1799}
J. Van der Jeugt, {\it J. Math. Phys.} {\bf 34}, 1799-1806 (1993).

\bibitem{Quesne81}
C. Quesne, {\it Phys. Lett. B} {\bf 304}, 81-88 (1993).

\bibitem{JPA29}
P. P. Raychev, R. P. Roussev, P. A. Terziev, D. Bonatsos and N. Lo Iudice,
{\it J. Phys. A: Math. Gen.} {\bf 29}, 6939-6949 (1996).

\bibitem{STK593}
Yu. F. Smirnov, V. N. Tolstoy and Yu. I. Kharitonov, {\it Yad. Fiz.} {\bf 53},
959-980 (1991) [{\it Sov. J. Nucl. Phys.} {\bf 53}, 593-605 (1991)].

\bibitem{STK690}
Yu. F. Smirnov, V. N. Tolstoy and Yu. I. Kharitonov, {\it Yad. Fiz.} {\bf 56},
223-244 (1993) [{\it Phys. At. Nucl.} {\bf 56}, 690-700 (1993)].

\bibitem{Martin209}
T. P. Martin, T. Bergmann, H. G\"ohlich and T. Lange, {\it Chem. Phys. Lett.}
{\bf 172}, 209-213 (1990).

\bibitem{Clem44}
K. Clemenger, {\it Phys. Rev. B} {\bf 44}, 12991-13001 (1991).

\bibitem{Knight2141}
W. D. Knight, K. Clemenger, W. A. de Heer, W. A. Saunders, M. Y. Chou and
M. L. Cohen, {\it Phys. Rev. Lett.} {\bf 52}, 2141-2143 (1984).

\bibitem{Balian76}
R. Balian and C. Bloch, {\it Ann. Phys. (N.Y.)} {\bf 69}, 76-160 (1972).

\bibitem{Gutz}
M. C. Gutzwiller, {\it J. Math. Phys.} {\bf 12}, 343-358 (1971).

\bibitem{Brec2271}
C. Br\'echignac, Ph. Cahuzac, F. Carlier, M. de Frutos and J. Ph. Roux,
{\it Phys. Rev. B} {\bf 47}, 2271-2277 (1993).

\bibitem{Morse}
P. Morse, {\it Phys. Rev.} {\bf 34}, 57-64 (1929).

\bibitem{Nilsson1}
S. G. Nilsson, {\it Mat. Fys. Medd. K. Dan. Vidensk. Selsk.} {\bf 29}, 16
(1955).

\bibitem{Nilsson2}
S. G. Nilsson and I. Ragnarsson, {\it Shapes and Shells in Nuclear
Structure} (Cambridge University Press, Cambridge, 1995).

\bibitem{Lerme2818}
J. Lerm\'e, M. Pellarin, J. L. Vialle, B. Baguenard and M. Broyer,
{\it Phys. Rev. Lett.} {\bf 68}, 2818-2821 (1992).

\bibitem{Persson}
J. L. Persson, R. L. Whetten, H. P. Cheng and R. S. Berry, {\it Chem. Phys.
Lett.} {\bf 186}, 215-222 (1991).

\bibitem{Nishi}
H. Nishioka, K. Hansen and B. R. Mottelson, {\it Phys. Rev. B} {\bf 42},
9377-9386 (1990).

\bibitem{Nishi2}
H. Nishioka, {\it Z. Phys. D} {\bf 19}, 19-23 (1991).

\bibitem{Genzken}
O. Genzken, M. Brack, E. Chabanat and J. Meyer, {\it Ber. Bunsenges. Phys.
Chem.} {\bf 96}, 1217-1220 (1992).

\bibitem{Pell96}
M. Pellarin, J. Lerm\'e, B. Baguenard, M. Broyer and J. L. Vialle,
{\it Ber. Bunsenges. Phys. Chem.} {\bf 96}, 1212-1215 (1992).

\bibitem{Pell98}
M. Pellarin, B. Baguenard, M. Broyer, J. Lerm\'e and J. L. Vialle,
{\it J. Chem. Phys.} {\bf 98}, 944-950 (1993).

\bibitem{VMI}
M. A. J. Mariscotti, G. Scharff-Goldhaber and B. Buck, {\it Phys. Rev.}
{\bf 178}, 1864-1887 (1969).

\bibitem{Flugge}
S. Fl\"ugge, {\it Practical Quantum Mechanics} (Springer, Berlin, 1974).

\bibitem{Cooper112}
I. L. Cooper, {\it Chem. Phys.} {\bf 112}, 67-75  (1987).

\bibitem{Cooper25}
I. L. Cooper, {\it J. Phys. A: Math. Gen.} {\bf 25}, 1671-1683 (1992).

\end{thebibliography}
\end{document}